\def\aa{{A\&A}}

\def\aj{{AJ}}
\def\annrev{{ARA\&A}}
\def\apj{{ApJ}}
\def\apjs{{ApJS}}

\def\mnras{{MNRAS}}

\def\deg{$^{\circ}$}

\def\farcs{\hbox{$.\mkern-4mu^{\prime\prime}$}}

\documentclass[cup5b]{caps}
\usepackage{graphicx}
\usepackage{amssymb}
\usepackage{ociwsymp1}   

\HeadText{C. M. Carollo}

\def\lta{{\>\rlap{\raise2pt\hbox{$<$}}\lower3pt\hbox{$\sim$}\>}}
\def\gta{{\>\rlap{\raise2pt\hbox{$>$}}\lower3pt\hbox{$\sim$}\>}}

\begin{document}

\pagenumbering{arabic}

\author[]{C. MARCELLA CAROLLO\\ Eidgen\"ossische Technische Hochschule, CH-8093
  Zurich, Switzerland}
 
\chapter{The Inner Properties of \\ Late-type Galaxies}

\begin{abstract}

I review some recent results on the inner properties of disk galaxies,
and highlight some issues that require either observational or
theoretical clarification and that are important for constructing a
consistent picture of the formation of the local disk galaxy
population.

\end{abstract}

\section{Disk Galaxies: Recipe Still Missing}

We have not yet achieved a self-consistent theoretical scenario for
the formation of the local disk galaxy population, a population which,
in crude terms, is a mix of ``bulged'' and ``bulgeless'' disks.  The
origin of the bulges is vigorously debated.  In a recent paper in the
proceedings of the 1998 workshop on ``The Formation of Bulges''
(Carollo, Ferguson, \& Wyse 1999), Renzini (1999) voices with emphasis
the long-standing belief that bulges are nothing more or less than small
elliptical galaxies.  Basing his argument on the similarity between
the magnesium line strength Mg$_2$ versus absolute $r$ magnitude $M_r$
(and magnesium line strength Mg$_2$ and velocity dispersion $\sigma$;
Jablonka, Martin, \& Arimoto 1996) relations for bulges and ellipticals, he
writes: {\sl ``...The close similarity of the Mg$_2-M_r$ relations
argues for spiral bulges and ellipticals sharing a similar star
formation history and chemical enrichment.  One may argue that origin
and evolution have been very different, but differences in age
distribution are precisely compensated by differences in the
metallicity distributions. This may be difficult to disprove, and I
tend to reject this alternative on aesthetic grounds. It requires an
unattractive cosmic conspiracy, and I would rather leave to others the
burden of defending such a scenario. In conclusion, it appears
legitimate to look at bulges as ellipticals that happen to have a
prominent disk around them, or to ellipticals as bulges that for some
reason have missed the opportunity to acquire or maintain a prominent
disk.''}

And yet, there is plenty of evidence from observations and numerical
experiments that bulges of spiral galaxies may differ significantly from
elliptical galaxies.  In a pioneering paper, Kormendy (1993; see also
Kormendy, Bender, \& Bower 2002) reports, for some Sb bulges, $V/\sigma$ values
that are above the oblate line describing the isotropic spheroids in the
$V/\sigma$-$\epsilon$ diagram (with $V$ the maximum velocity, $\sigma$ the
mean velocity dispersion, and $\epsilon$ the mean ellipticity of the spheroid;
Binney \& Tremaine 1987), and makes the point that at least some of the dense
structures that are seen inside the disks may actually themselves be
disklike systems (``pseudo-bulges'').  In numerical simulations,
three-dimensional stellar structures result from secular evolution processes
that are driven by dynamical instabilities inside the preexisting disks.  The
fire-hose (or buckling) instability that is seen in simulations of stellar
disks can scatter the stars originally in a stellar bar above the plane of the
disk, into what resembles a bulgelike structure (Raha et al.\ 1991). A
stellar bar can also drive a high inflow rate of gas toward the center of the
disk (Shlosman, Frank, \& Begelman 1989); if a mass concentration of the order
of $\sim 1\%$ of the total mass is accumulated in the center, this can
disrupt the regular orbits supporting the bar and again scatter the stellar
orbits above the plane of the disk (Pfenniger \& Norman 1990; Norman, Sellwood, \& Hasan 1996).

The disks of spiral galaxies also elude us.  On large galactic scales,
they are a rather homogeneous family, as indicated, for example, by their light
profiles, which appear to be exponential over several disk scale lengths (de
Jong 1995), the rather common asymptotically flat rotation curves (Persic \&
Salucci 1995), and the Tully-Fisher relation, which holds over a broad range of
surface brightness and mass (Strauss \& Willick 1995).  On smaller scales,
however, where they physically overlap with the bulges (and the rest of the
inner structure), disks are not well understood, either observationally or
theoretically.  Within hierarchical formation schemes, the standard recipe to
explain the formation of disks contains three key elements: (1) 
the angular momentum originates from cosmological torques (Hoyle 1953), 
(2) the gas and dark matter within virialized systems have initial angular
momentum distributions that are identical (Fall \& Efstathiou 1980), and (3)
the gas conserves its specific angular momentum when cooling (Mestel
1963).  These rules are routinely assumed in the (semi-)analytical
descriptions of disk galaxies.  In contrast, the highest resolution
cosmological simulations that include both baryons and cold dark matter (CDM)
find significant angular momentum loss for the baryons, especially in the
central few kpcs of galaxies (Steinmetz \& Navarro 1999).  Furthermore, even
when disks are assumed to form smoothly and conserving their angular momentum,
the resulting disks are more centrally concentrated than single-exponential
structures (Bullock et al.\ 2001; van~den~Bosch 2001; van~den~Bosch et al.\ 
2002).  Disks with high central densities are seen in the highest-resolution
CDM simulations, in which the resulting galaxies have realistic sizes, but a
region with low angular momentum and high density is always present at the
center (e.g., Governato et al.\ 2003). It is still a matter of debate whether
this is a feature of the structure formation model or is indicative of the
lack of a proper treatment of physics (e.g. the effect of supernovae feedback;
Springel \& Hernquist 2002).  Although the CDM simulations still have room for
improvement, they could well be correct in their prediction that the central
parts of disks might really have quite low angular momentum and high
concentration as a result of formation.  Although warm dark matter alleviates 
the angular momentum ``catastrophe'' (i.e., the loss of angular momentum by the 
baryons), the angular momentum distributions of warm dark matter halos is
identical to that of CDM halos (Knebe, Islam, \& Silk 2001; Bullock, Kravtsov, 
\& Colin 2002).  Therefore, these halos also predict an excess of low-angular 
momentum material.

Clearly the central regions of disk galaxies hold important clues to understanding
fundamental issues of galaxy formation. Shaping a consistent theory of bulge
and disk assembly requires a better understanding of nearby disk galaxies on
the nuclear and circumnuclear scales. High-resolution studies of real and
simulated disk galaxies are still in their infancy, but have made their first
steps in the last few years.  Recent reviews on disk galaxies and their
subcomponents are presented by Wyse, Gilmore, \& Franx (1997) and Carollo et
al.\ (1999).  In this review I focus on some recent developments on the
central regions of nearby disk galaxies, and discuss some of the related
important issues that require future attention.  In order to remain faithful
to the original studies, and following customary classification schemes, I
will often discuss the results maintaining a distinction among systems of 
early, intermediate, and late types; however, it is this very distinction that
I challenge in my concluding remarks.

\section{First Fact: Complexity is the Rule}

All recent high-resolution studies consistently report a large complexity in
the inner regions of at least half of the local disk galaxy population of all
Hubble types. On scales smaller that 1 kpc, more than half of the galaxies
host inner bars (within bars), dust or stellar or gaseous disks, spiral-like
dust lanes, star-forming rings, spiral arms, a central cluster (\S~1.5), or 
simply irregular central emission.  (e.g., Carollo et al.\ 1997a; Carollo, 
Stiavelli, \& Mack 1998; Martini \& Pogge 1999; Laine et al.  2002; B\"oker, 
Stanek, \& van~der~Marel 2003).

Nuclear bars have received particular attention, as they are claimed to play
an important role in feeding gas into the centers of galaxies (Shlosman et
al.\ 1989), potentially building central nuclei and bulges, and fueling
nuclear activity.  Intimate links between bars and central starbursts, in
particular circumnuclear star-forming rings, are supported by observations
(Knapen, P\'erez-Ram\'\i rez, \& Laine 2002).  Other studies point out, 
however, that, on the nuclear scales, stellar rings and inner disks inside 
large-scale bars of moderately inclined galaxies could be mistaken for 
secondary bars or even coexist with them, producing erroneous statistics for 
the occurrence of nuclear bars (Erwin \& Sparke 1999).  Still, {\it bona 
fide}\ secondary inner bars, typically about 250 pc--1 kpc in size ($\sim$12\% 
the size of their primary bars) appear to be present in as many as 40\% of all 
barred S0--Sa galaxies (Erwin \&
Sparke 1999). Larger samples of early-type galaxies confirm a high frequency
of detection of bars-within-bars (Rest et al.\ 2001).  This high frequency is
interpreted to indicate that secondary nuclear bars are relatively long-lived
structures. The presence or absence of secondary bars appears to have no
significant effect on nuclear activity (as previously reported by, e.g., Regan
\& Mulchaey 1999). In contrast, nuclear spirals, dusty or star-forming nuclear
rings, and off-plane dust are reported to be very often accompanied by LINER
or Seyfert nuclei (Erwin \& Sparke 1999; Martini \& Pogge 1999).

Circumnuclear starburst rings have also been thoroughly investigated (e.g.,
Maoz et al.\ 1996, 2001). These rings appear to be a common mode of starbursts
in relatively early-type disk galaxies, and are thought to be associated with
inner-Linblad resonances. They contain super star clusters with total
luminosities as high as $M_V \approx -15$ mag ($L_V \approx 1.3\times10^8
L_{V,\odot}$), radii of the order of a few parsecs, and masses in excess of
$10^4 M_\odot$. These clusters are very similar to the super star clusters
formed in merging systems (e.g., Whitmore et al.  1999; Hunter et al. 2000):
they are bound systems, believed to evolve into stellar structures similar to
globular clusters.  The starburst rings are thought to be likely associated
with bar-driven inflow.  Schinnerer et al.\ (2002) report in the double-barred
galaxy NGC 4303, which also hosts a circumnuclear star-forming ring (Colina 
\& Arribas 1999), an extremely good agreement between the observed overall gas 
geometry and dynamical models for the gas flow in barred galaxies
(Englmaier \& Shlosman 2000).  Observational evidence seems thus to be
accumulating in support of the theoretical prediction that disk instabilities
on large scales are major drivers of evolution on circumnuclear (and
nuclear) galactic scales.  Similar to nuclear bars, circumnuclear star-forming 
rings are found to coexist with AGNs but are not associated one-to-one 
with AGN activity.

\section{News on Bulges}

It has been known for some time that many bulges have a radial light profile
that is not an elliptical-like $r^{1/4}$ law (Andredakis \& Sanders 1994; de
Jong 1995; Courteau, de~Jong, \& Broeils 1996); instead, they are reasonably
well described by an exponential light profile.  Incidentally, the bulge of
our own Milky Way also has an exponential light profile (Binney, 
Gerhard, \& Spergel 1997).
Recent high-resolution investigations using data from the {\it Hubble Space
Telecope (HST)}\ have strengthened the evidence for exponential light profiles
down to the smallest scales at the end of the spheroids luminosity sequence
(Carollo et al.\ 1998, 2001; Carollo 1999).  Several studies
have used the generalized surface density profile $I(r) \propto
{\rm exp}[-(r/r_o)^{(1/n)}]$ introduced by S\'ersic (1968) to model the bulge light
(Andredakis, Peletier, \& Balcells 1995; Graham 2001; MacArthur, Courteau, \& Holtzman 2003).
These studies report shape-parameter values for bulges of late-type spirals
ranging between $n$ = 0.1 and 2.  Some of the S\'ersic analyses attribute a
significant meaning to the derived precise values of $n$ (Graham 2001; Balcells
et al.\ 2003). Tests based on simulated data, however, show a large dependence
of the derived parameters on, for example, the input parameters; indeed, 
MacArthur et al. (2003) stress that, on average, the underlying surface 
density profile for the late-type bulges is adequately described by an 
exponential distribution.  The same studies show the existence of a coupling 
between bulges and disks that is manifested by an almost-constant scale 
lengths ratio $h_{\rm bulge}$/$h_{\rm disk} \approx 0.1$ for late-type spirals, 
and a similar scaling relation even for earlier-type systems. This is 
interpreted to indicate a similar origin for bulges of all sizes in hosts 
of any Hubble type.

For more massive, early-type bulges, ground-based studies using the
S\'ersic law to describe their light distribution have found values of
$n$ close to, or even in excess of the elliptical-like de~Vaucouleur's (1948)
value of $n=4$ (Graham 2001). However, the analysis of high-resolution
{\it HST} images for a sample of early-type bulges provides values of the
S\'ersic shape index $n$ not in excess of $\sim 3$ (Balcells et al.\
2003). The difference in the estimates for $n$ is due to the
contribution of photometrically distinct central point sources, which
at ground-based resolution are confused for bulge light (see \S 1.5).
Balcells et al. interpret the $n<3$ S\'ersic shape indices
in the massive bulges as an indication that even these systems, like
the smaller exponential-type bulges, are not the outcome of violent
relaxation during collisionless accretion of matter.  Both in the
ground-based and {\it HST} analyses, a trend remains between the bulge
S\'ersic shape parameter $n$ and the bulge luminosity and
half-light radius; the trend is in the direction of brighter, bigger
bulges having larger $n$ values (Graham et al. 2001; Balcells et al.\
2003).

Detailed studies of the integrated stellar populations of bulges of all Hubble
types have also been pushed forward by the availability of high-resolution
multi-color images from the {\it HST} (Peletier et al.\ 1999; Carollo et al.\ 2001).
The independent analyses agree on the basic result that (1) massive
early-type bulges have very red colors, unambiguously indicating old ages
($\gta 8$ Gyr) for the average stellar populations of these systems, and 
(2) the smaller, later-type (almost) exponential bulges have on average
significantly bluer colors.

Kinematically, the Sb pseudo-bulges studied by Kormendy (1993) represent the
extreme case of a general behavior shown by bulges of any Hubble type and
mass: these all appear to have kinematic properties that are closer to
disklike structures rather than to elliptical galaxies.  Indeed, based on the
comparison between minor axis radial velocity dispersions of disks and bulges,
Falc\'on-Barroso et al.\ (2003) report that even the early-type, massive bulges
are actually thickened disks.

The following important considerations on bulges that emerge from the above 
analyses. (1)
The earlier-type bulges form a continuum with the late-type bulges
in terms of the shapes of the surface brightness profiles.  The smallest bulges
are exponential structures, and the largest appear to be intermediate cases
between the exponential and the elliptical-galaxy ones.  (2) Bulges of
spirals are coupled to their host disks in a similar way along the Hubble
sequence (with a possible weak trend toward marginally higher
$h_{\rm bulge}$/$h_{\rm disk}$ ratios for early-type bulges).  (3) There is a
spread in average stellar metallicity and ages amongst bulges, but also a
clear trend toward smaller bulges being less enriched and younger stellar
structures than the more massive, earlier-type bulges.  (4) Bulges of any
size show some kinematic features that are typical of disks.

\section{News on Disks}

Recent surveys with the {\it HST} that have focused on the late-type, allegedly
bulgeless Scd--Sm disks, find that only $\sim$30\% of these systems have light
profiles consistent with being single-exponential structures (B\"oker et al.\ 
2003). The remaining disks are not well fitted by a single exponential; in
particular, the surface brightness in the central few kiloparsecs exceeds the
inward extrapolation of the outer exponential disk.  The surface brightness
profiles of many of these late-type disks are often equally well described
either by the sum of two exponential components, or by a single S\'ersic profile
over the entire radial range with shape parameter $n$ up to a value of 2.5.
In the earlier-type systems, a second central exponential component in
addition to the outer exponential disk is typically interpreted as a bulge
component.  B\"oker et al.\ (2003) suggest that the frequent detection of such
central exponential ``excesses'' also in systems that, according to the
classical classification scheme, should host no bulge component, together with
the fact that a single S\'ersic profile is often a good alternative to the sum
of two exponentials, may indicate that in fact these excesses are not bulges,
but rather denser regions of the disks themselves.

A key issue in the context of understanding the nature of the central regions
of disk galaxies is one of definitions (see Carollo et al.\ 1999). B\"oker et
al.\ distinguish between what they call ``the modern theorist'' view, assumed to
be the correct one, which asserts that a bulge is a kinematically hot component
with an extended three-dimensional structure, and the ``observers'' view, which,
in photometric studies, relies on the assumption that disks are exponential
structures and that bulges are identifiable as additional light (mass)
contributions in the central regions.  All the photometric analyses of the
local (and distant; see, e.g., Shade et al.\ 1996) disk galaxy population that
are aimed at studying bulge properties indeed assume a constant scale length
exponential profile for the disk, and attribute to a bulge any central
concentration of light in excess of the inward extrapolation of the outer,
constant scale length disk (e.g., Andredakis \& Sanders 1994; Andredakis et
al.\ 1995; de~Jong 1995; Courteau et al.\ 1996; Carollo et al.\ 1998;
Balcells et al.\ 2003; MacArthur et al.\ 2003).  B\"oker et al. (2003)
mention a lack of theoretical support for the disks being exponential, and
point out that the assumption that disks remain exponential all the way into
the center may not be correct, i.e., that the operational definition of bulges
adopted in photometric studies may lead to attributing to a bulge what
actually belongs to the disk. It is certainly not an easy task to disentangle
into distinct subcomponents the centers of galaxies, where all of these
subcomponents are expected to reach their largest densities.

\section{A Zoom on the Centers: Point Sources and Distinct Nuclei}

Photometrically distinct compact nuclei have been known for a while to reside
in the centers of bulgeless, weakly active or inactive late-type disks (e.g.,
Kormendy \& McClure 1993; Matthews \& Gallagher 1997).  Extensive surveys with
the {\it HST} show, however, that distinct pointlike or compact nuclei are the 
rule rather than the exception in the centers of all sorts of disk galaxies.

An optical ($V$, WFPC2) and near-infrared ($H, J$, NICMOS) survey of $\sim
100$ intermediate-type, Sb-Sc spirals shows that $\sim 70\%$ of these
systems host distinct nuclei with visual absolute magnitudes $-8 \gta M_V \gta
-16$, comparable at the bright end with young super star clusters in
starbursting galaxies (Carollo et al.\ 1997a, 1998, 2002).  Some of 
these nuclei appear pointlike in the {\it HST} images.
However, many are marginally resolved with half-light radii
$\sim$0\farcs1--0\farcs2, corresponding to linear scales of a few to up to
$\sim$20 pc. These nuclei cover a large range of colors in the range 
$-0.5\,{\rm mag}\, \lta V-H \lta\, 3\,{\rm mag}$.  Statistically, the 
distribution of $V-H$ colors is broader for the nuclei than for the 
surrounding galactic structure; this
suggests that star formation, AGN activity, dust reddening, or a
combination of these is generally present in the nuclei embedded in
intermediate-type hosts. Most of the nuclei, at {\it HST} resolution, are 
located at the galaxy centers and appear to be round, star-cluster-like, 
structures.
However, in some cases the nuclei are offset from the isophotal/dynamical
centers of the host galaxies or show some degree of elongation (as it is the
case for, e.g., the nucleus of M33; Kormendy \& McClure 1993; Lauer et al.
1998; Matthews et al. 1999). Both the displacement, which is typically of a
few tens of parsecs, and the elongation appear to be uncorrelated with either
the luminosity and color of the nucleus, or with the galaxy type. Searches for
trends with other galactic subcomponents reveal no clear relationship between
the distinct nuclei and nuclear (or even larger-scale) bars: bars are neither
ubiquitous nor unusual in nucleated intermediate-type spirals.  Some of the
nuclei are embedded in exponential-type bulges; actually, every such bulge in
this sample hosts a distinct nucleus in its isophotal center or slightly
offset from it.  These nuclei have low to moderate luminosities, 
$-8\, {\rm mag}\, \gta\, M_V\, \gta\, -12$ mag.  Selection effects may be 
present.  Brighter nuclei are
typically embedded in very complex circumnuclear structures that do not
allow for the derivation of reliable bulge parameters.  Moreover, 
exponential-type bulges lying under substantial circumnuclear structure but
hosting no central nucleus would also, for similar reasons, drop from the
sample. The nuclei that have been identified inside the exponential-type 
bulges have colors compatible with those arising from stellar populations. 
Under the assumption of no dust reddening, average stellar ages of $\sim$1 Gyr 
are inferred for these relatively faint central star clusters; these ages,
in turn, imply masses of about 10$^6$ to 10$^8$ $M_\odot$.

Studies of large samples of Scd and later-type disks also find compact,
distinct nuclei, identified as ``star clusters,'' in about 75\% of the
population (B\"oker et al.\ 2002). Their distribution of absolute luminosities
has a FWHM of about 4 mag and a median value of $M_I = -11.5$ mag. These
luminosities, as well as the sizes of these star clusters, are comparable to
those of the nuclei embedded in the earlier-type galaxies.  For 10 nuclei in
the B\"oker et al.\ (2002) sample, Walcher et al.\ (2003) report spectroscopic 
estimates for the stellar ages that are smaller than about 1 Gyr, and
masses of the nuclei estimated from stellar velocity dispersions in the range
10$^6-10^8$ $M_\odot$. These masses are similar to the photometry-based
mass estimates derived for the clusters embedded in the exponential-type
bulges, and are one order of magnitude higher than expected from stellar
synthesis models.  These authors interpret this discrepancy as due to old(er)
stellar generations contributing to the total mass of the nuclei. They
conclude that the nuclei of late-type disks are grown in multiple star
formation events, and suggest that this mechanism may contribute to the
formation of Kormendy's pseudo-bulges.

At the other extreme of the Hubble sequence, the large majority of the
early-type spiral galaxies that host massive bulges also require a central
component, in addition to the bulge and the disk, to reproduce the observed
light profiles (Balcells et al.\ 2003).  In most cases, this additional
component appears to be a point source at the resolution of NICMOS on the 
{\it HST}.  Balcells et al. point out that in ground-based data these
additional light contributions cannot be disentangled from the bulge light
and may produce overestimates of the derived S\'ersic $n$ parameter 
(see \S 1.3).

\begin{figure*}[t]
\vspace{1.0cm}
\includegraphics[width=1.00\columnwidth,angle=0]{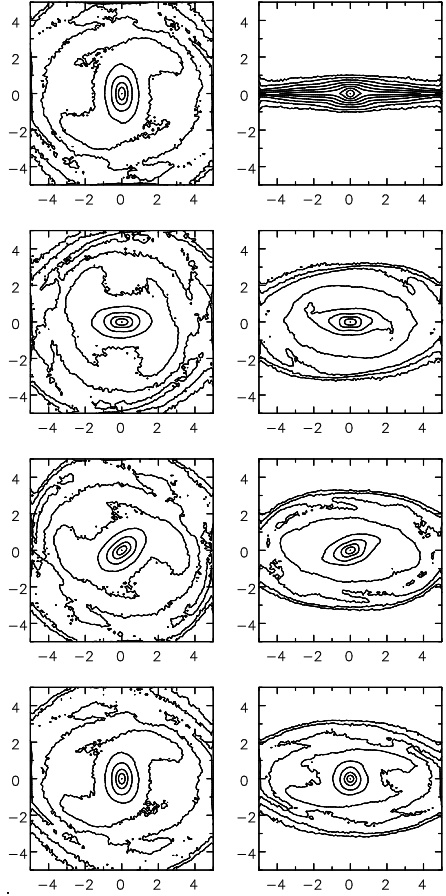}
\vskip 0pt \caption{
  Contour plots for a live-disk/frozen-halo simulation after
  the development and buckling of a stellar bar.
  The halo has a Hernquist (1990) profile; the disk initially has a
  single-exponential profile. Shown are the face-on and edge-on views of the
  system (upper panels, left and right, respectively), and two different views
  with disk inclination angles of 30\deg\ (left) and 60\deg\ (right),
  respectively.  Different panels from second-top to bottom show different
  orientation angles for the buckled bar of 0\deg, 45\deg, and 90\deg,
  respectively, with respect to the major axis of the disk. Axes are in units
  of the initial exponential disk scale length. (From Debattista et al.\ 2003.)
\label{Figure 1.1}}
\end{figure*}

\begin{figure*}[t]
\vspace{-2.9cm}
\includegraphics[width=1.00\columnwidth,angle=0]{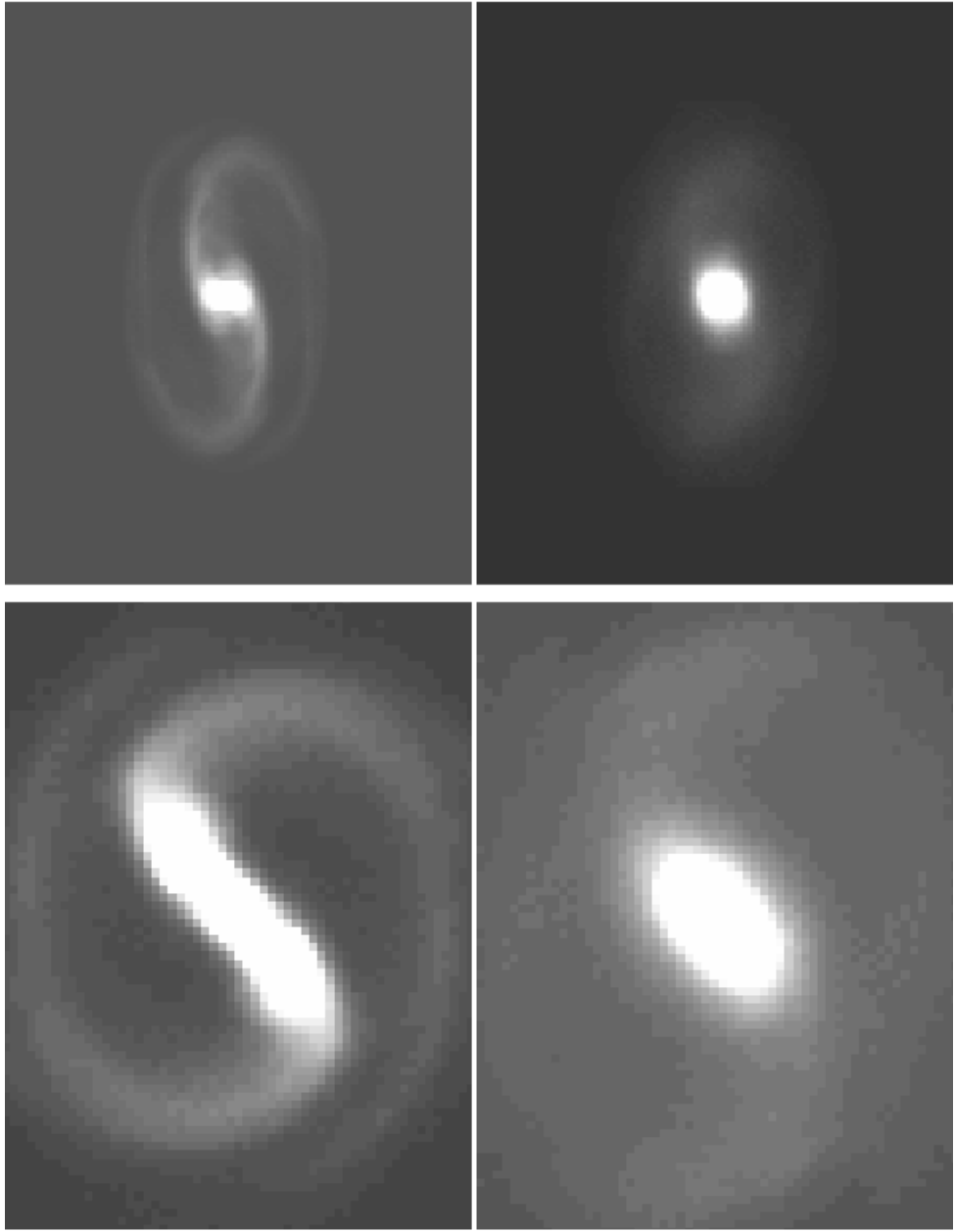}
\vskip 0pt \caption{
  A Hernquist halo simulation (upper panels) and a core halo simulation
  (lower panels). The images are taken before the buckling (left panels) and
  after the buckling (right panels) of the bars. The models have inclination
  angles of 60\deg\ and 30\deg, and bar orientation angles of 90\deg\ and
  45\deg, respectively. (From Debattista et al.\ 2003.)
\label{Figure 1.2}}
\end{figure*}

\begin{figure*}[t]
\vspace{-5.3cm}
\hspace{+0.5cm}
\includegraphics[width=0.95\columnwidth,angle=0]{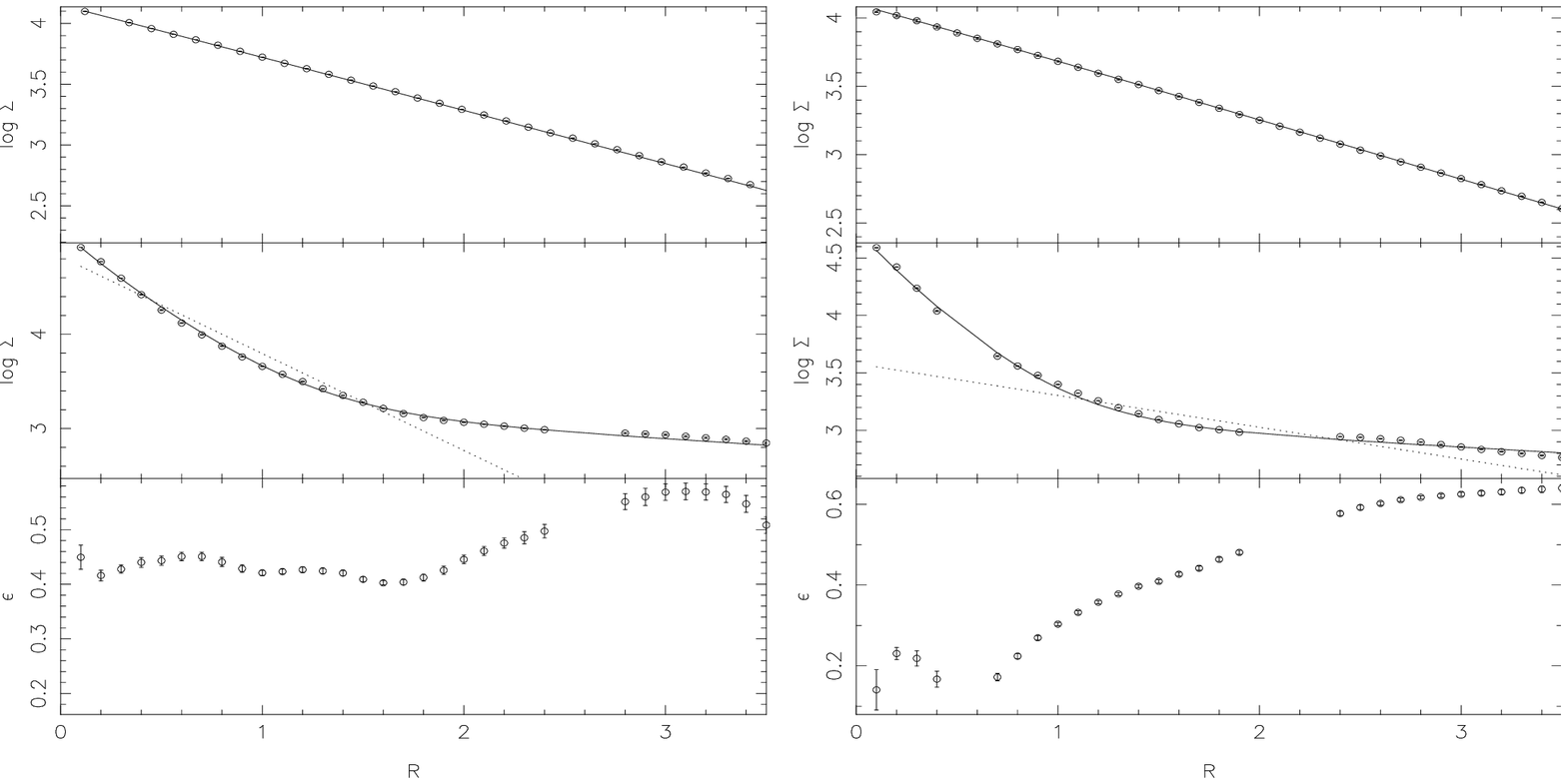}
\vskip 0pt \caption{
  For the same core halo (left) and Hernquist halo (right) models shown
  in Figure 1.2, we plot here the initial surface density profiles
  (upper panels: superimposed with the single-exponential fits), the surface
  density profiles after buckling (middle panels: dotted lines, the single
  exponential fits; solid lines, exponential plus S\'ersic fits), and the
  ellipticity profiles after buckling (lower panels). The x-axis is in units
  of the initial exponential disk scale length. (From Debattista et al.\ 2003.)
\label{Figure 1.3}}
\end{figure*}

\section{Future Challenges}

\subsection{Disks}

The claim that there is no strong theoretical support for the disks to be
exponential (B\"oker et al.\ 2003) is partly substantiated by the fact that
theories are often tuned so as to reproduce an exponential light profile.
While reproducing exponential light profiles is considered to be a test for
viable models of disk formation (e.g., Dalcanton, Spergel, \& Summers 1997;
Silk 2001), it is certainly true that most of the current disk models and
simulations do predict large mass fractions with low-angular momentum material
that is in excess of the extrapolation of the outer exponential density
distribution to the center (see \S 1.1 and references therein).  However, the
real problem is to understand whether this excess low-angular momentum material
does remain in the disk, or rather forms a three-dimensional bulge.

Furthermore, it is also possible that, rather than being born denser
than exponential, the disks may become so during the subsequent galactic
evolution. Indeed many processes can occur during the Hubble time that can
transform exponential disks into the more complex structures that are observed
in late-type disks. In this context, it is worth stressing that a third of the
late-type disks remais well described by the simple single-exponential form,
showing that this channel of disk formation is indeed accessible to real
galaxies.

An obvious example of change in central concentration in disks is the one
induced by the formation and subsequent buckling of a stellar bar.  To study
at unprecedented resolution the effects of secular evolution processes on the
central regions of disks, we are conducting state-of-the-art $N$-body and
$N$-body+SPH numerical simulations of disk galaxies (details will be published 
in Debattista et al. 2003 and Mayer et al. 2003).  The first $N$-body 
experiments that I briefly discuss here (from Debattista et al.\ 2003) consist 
of live disk components inside frozen halos described either by a spherical
logarithmic potential with a central core or a cuspy Hernquist (1990)
potential. The initially axisymmetric disks are modeled assuming an
exponential profile with a Gaussian thickening; the disks are represented by
$(4 - 7.5)\times10^6$ equal-mass particles. The spatial resolution that is
achieved in the central regions is $\sim$50 pc. The simulations are run on a
3-D cylindrical polar-grid code (described in Sellwood \& Valluri 1997).  In
certain areas of parameter space, the axisymmetric systems are found to be
unstable and form bars.  Systems in which bars fail to form have only modest
heating, indicating that our results are not driven by noise.  Every 20
time steps of the disk evolution, we measure the disk velocity dispersions and
streaming velocities in annuli, the amplitude of the bar from the $m=2$
Fourier moment, and the amplitude of the buckling from the $m=2$ Fourier
moment of the vertical displacement of particles.  We use these quantities to
determine when the bar forms, when it buckles, and the evolution of disk
properties such as mass density, morphological diagnostics (for any
inclination angle of the disk and orientation of the bar), and the $V/\sigma$
ratio.  As an example, Figure 1.1 shows, for various disk inclination angles
and tilt angles of the bar with respect to the disk major axis, the isophotal
contours of the projected surface density of the system after the bar has
formed and buckled. As already pointed out in, for example, Raha et al.\ 
(1991), for some projections, including but not uniquely for the edge-on one, 
the buckled bar looks very much like a normal (three-dimensional)
``rounder-than-the-disk'' bulge.  Figure 1.2 shows, for two different
simulations, how the systems would appear on the sky as observed from a
specific line-of-sight before and after the buckling of the bars.  For the
same two models, Figure 1.3 plots the initial surface density profiles
(exponential by construction), the surface density profiles after the buckling
of the bars, and the post-buckling ellipticity profiles of the systems. Figure
1.3 points out that an initially exponential disk that ``nature'' makes 
can be observed to
be a more centrally concentrated structure at a later stage, after it has
formed and buckled a stellar bar.  The final, post-buckling profile in the
simulations is well described by the sum of an outer exponential disk and an
inner S\'ersic component, as observed in real disk galaxies.  If the
denser-than-exponential profiles in the real late-type spirals were due to the
effects of nurture rather than nature, the inner light/mass excesses in the
late-type disks would be better associated with ``bulge'' structures; 
that is, they would be the ``bulges'' produced by secular evolution of the 
disks, which has been extensively discussed in the literature.

Bars are not the only possible solution to increase the central densities in
disk galaxies by means of processes that occur after the original baryonic
collapse inside the dark halos: viscosity may be important (e.g., Lin \&
Pringle 1987), and also mergers, satellite accretion, dynamical friction of
globulars, etc.  Nonetheless, it is fair to conclude that, at this stage, the
issue whether the disks are born as denser-than-exponential structures remains
open. If this were the case, it will be important to quantify the systematic
uncertainties on, for example, bulge scale lengths and luminosities, black hole
masses and other galactic properties that are derived assuming that nature,
when it makes a disk, makes it exponential.

\subsection{Bulges}

The investigations of the past few years indicate that even the most massive,
early-type bulges are not $r^{1/4}$-law systems and have disklike imprints in
their kinematics.  How do we reconcile, under a common denominator, the 
differences between bulges and ellipticals with the quoted similarities of 
stellar population and scaling laws? It is certainly not clear what, for 
instance, the Mg$_2$ index and the velocity dispersion
$\sigma$ represent in the Mg$_2-\sigma$ relation.  Are the key 
parameters metallicity, age, or a combination of the two?  Are they the 
depth of potential well, local physics of star formation, or, again, a 
combination of the two?  Local physics imposes
thresholds for star formation (e.g., Meurer et al.\ 1997), which is likely to
have an impact on scaling laws such as the Mg$_2-\sigma$ relation.  Indeed,
the same Mg$_2-\sigma$ relation is observed to hold over orders of magnitude
of scale lengths, in systems that are very different, ranging from
elliptical galaxies to dwarf spheroidals (Bender, Burstein, \& Faber 1993).
The conclusion is that the Mg$_2-\sigma$ and similar relations are certainly
telling us something important about the formation of stellar systems over a
large range of scales, but not necessarily that they all share a similar
formation process.

On the other hand, the claims that violent relaxation is not a major player in
the formation of bulges, based on the observed S\'ersic profiles with $n\lta3$
(Balcells et al.\ 2003) may also be premature.  The consequence of violent
relaxation during dissipationless processes such as stellar clumpy collapses
(van~Albada 1982), mergers of disk galaxies (Barnes 1988), satellite accretion
onto disk galaxies (Aguerri, Balcells, \& Peletier 2001) is to produce an
$r^{1/4}$ profile. However, other studies of violent relaxation in a 
finite volume
show deviations from the $r^{1/4}$ law (Hjorth \& Madsen 1995).  Furthermore,
the same problem of separating nature from nurture may be relevant also in
this context.  Physical processes may occur during the Hubble time that modify
the stellar density profiles in the centers of galaxies, including dynamical
friction of globular clusters, dissipative accretion of matter,
black hole-driven cusp formation, mergers of black holes (quantitative studies
of the latter show that central mass deficits are created from the binding
energy liberated by the coalescence of the supermassive binary black holes;
see, e.g., Milosavljevi\'c et al.\ 2002, Ravindranath, Ho, \& Filippenko 
2003, and references therein).  Numerical studies of these processes are still 
rather sketchy and do not explore a vast volume of parameter space; 
nonetheless, they make the point that the nuclear stellar density profiles may 
be modified by subsequent evolution.  Quantitative work remains to be done to 
assess whether these or other processes can reproduce the $n \approx 3$ 
S\'ersic profiles typical of the massive bulges and the weak trend between 
S\'ersic shape parameter $n$ and bulge luminosity. The possibility that the 
disks may not be purely exponential also introduces additional uncertainties 
on the derived bulge parameters, including the shape index $n$. If the outer 
disk can have a S\'ersic shape with $n$ values as steep as
$\sim 2.5$, bulge-disk decompositions that use an exponential for the outer
disks can systematically offset the bulge parameters.  This could even open
the question as to whether the observed sequence in $n$ values between the 
late-type and early-type bulges is a pure bulge sequence, or, rather, at least 
in part a sequence of different underlying disk profiles.

\begin{figure*}[t]
\vspace{-6.8cm}
\hspace{-1.0cm}
\includegraphics[width=1.10\columnwidth,angle=0]{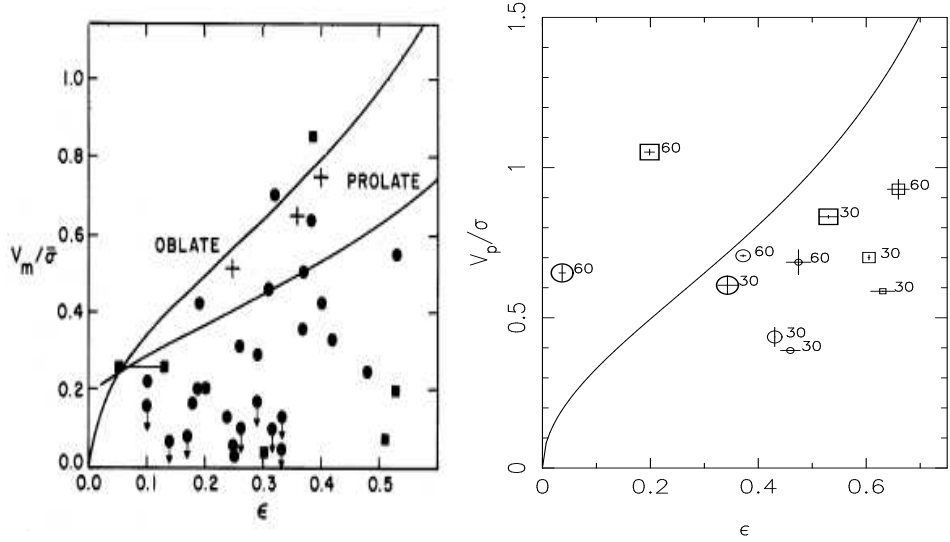}
\vskip 0pt \caption{
$V/\sigma$ versus ellipticity $\epsilon$ plane. The left panel reproduces
Figure 2 of Davies \& Illingworth (1983); points are the measurements for
observed spheroids, both bulges (crosses), and ellipticals (filled circles and
squares).  The right panel shows a similar plot derived from our simulations.
The oblate-rotator line is also plotted as a solid line.  The data points in
this panel refer to the end of the simulations, after the bars that have
formed in the disks have buckled.  The $V/\sigma$ and $\epsilon$ values for
the ``buckled bars'' are derived by averaging the relative profiles inside the
half-light radii of the inner S\'ersic components that are necessary, in
addition to the outer exponential disk components, to obtain good fits to the
surface density profiles after buckling.  Circles are used for the logarithmic
and squares for the Hernquist halo potentials. The size of the symbols refers
to the bar orientation angles (90\deg\ largest, 45\deg\ intermediate, 0\deg\
smallest). The disk inclination angle $i$ is explicitly indicated close to
  symbols.  From certain viewing angles and bars orientation angles, buckled
bars are indistinguishable from normal spheroids on the $V/\sigma$ versus
ellipticity $\epsilon$ plane. (From Debattista et al.\ 2003.)
\label{Figure 1.4}}
\end{figure*}

Concerning support for bulge-building secular evolution processes inside
preexisting disks, there is certainly at this point good evidence from
high-resolution numerical experiments that the intrinsic evolution of the
disks results in transformations of the disks, which can generate
three-dimensional structures that resemble bulgelike components.  Numerical
studies (Pfenniger \& Friedli 1991; Zhang \& Wyse 2000; Scannapieco \&
Tisseira 2003; Debattista et al.\ 2003, see Fig. 1.3) also show that the
bulgelike, three-dimensional structures that generally result from the
evolution of the disks have the rather low-$n$ S\'ersic profiles typical of real
bulges.  MacArthur et al.\ (2003) report that simulations by D. Pfenniger
(2002, private communication) of self-gravitating disks form bars that may
later dissolve into bulgelike components, which show a nearly universal ratio
of bulge-to-disk scale lengths, also in agreement with the observed
correlations. In the simulations, the universal ratio of bulge-to-disk
scale lengths is related to the stellar dynamics of the barred system, for 
example to the relative position of the vertical to horizontal resonances.  
There is
an additional important ingredient that has been missing so far in the debate
concerning the possibility that disk secular evolution processes play a
substantial role in forming bulgelike structures: namely, the bulges that
result from the secular evolution of the disks are, in contrast to what is
commonly asserted, not necessarily dynamically cold, ``disklike'' stellar
systems.  Due to the fact that eccentric orbits are quickly erased by shocks
(Friedli \& Benz 1995), the secular evolution of mostly gaseous disks indeed 
produces cold stellar structures such as the pseudo-bulges discussed by Kormendy
(1993); however, the buckling of stellar bars, for example, can produce
structures that are, at least from certain viewing angles, indistinguishable
from the alleged ``normal'' bulges in classical diagnostic planes such as the
$V/\sigma$-$\epsilon$ plane.  This is shown in Figure 1.4, where the locations
on the $V/\sigma$-$\epsilon$ of a few representative buckled bars from our
simulations are shown (right panel) in comparison with what is typically
considered the {\it bona fide}\ bulge behavior (left panel, figure from Davies 
\& Illingworth 1983).

In summary, from Figures 1.1--1.4, it is evident that, depending on the
viewing angle, buckled bars can appear as structures that are simultaneously 
rounder than the surrounding disks, photometrically identifiable as additional 
components in excess of outer exponential disks, and kinematically similar to 
what are considered to be ``{\it bona fide}\ bulges.'' In this light, it seems 
appropriate to question indeed what is a meaningful definition for 
a {\it bona fide}\ bulge.  Clearly, the situation is
more complex than what is captured in the theorist-versus-observer dichotomy
discussed by B\"oker et al.\ (\S 1.4).  First, from an observational
perspective, even early-type, {\it bona fide}\ bulges have been claimed to be
thickened disks (Falc\'on-Barroso et al.\ 2003). Second, from a theoretical
perspective, evolutionary disk processes such as the buckling of progenitor
bars inside the disks can produce structures that, in contrast to common
belief, are dynamically similar to the {\it bona fide}\ bulges that should be 
the benchmark for the comparison.  Thus, as with the photometric classification,
even the kinematic classification of bulges is quite fuzzy. Ultimately, this 
is due to the lack of a proper physical
boundary between structures that are forced into different categories by what
may be unfolding into an obsolete and confusing classification scheme.

\subsection{The Nature and Role of Nuclei and Central Black Holes}

Recent surveys show that central, distinct, compact components, in addition to
the disk and the bulge, are present in the large majority of disk galaxies of
all Hubble types.  Many are clearly star clusters with no AGN contamination.
This includes, for example, the ``naked'' ones in the late-type disks studied
by Walcher et al.\ (2003) and probably the relatively faint population of
nuclei embedded in the relatively clean surroundings of the exponential-type
bulges (Carollo et al.\ 1997a, 1998).  AGNs are known to be rare in late-type
galaxies (Ho, Filippenko, \& Sargent 1997; Ulvelstad \& Ho 2002; Ho 2003).  An
AGN component may, however, be present in a fraction of the nuclei.  This
would be statistically consistent with the fact that about 70\% of spirals
host a distinct nucleus, and about half of them are known to host some form of
AGN, even if weak (Ho et al. 1997).  Some of the point sources embedded in the
early-type bulges of Balcells et al.\ (2003) may also have an AGN origin or
component; pointlike sources associated with AGNs are seen in massive
elliptical galaxies (Carollo et al.\ 1997b,c; Ravindranath et al. 2001).

The young stellar ages plus high velocity dispersions of the central star
clusters of late-type disks reported by Walcher et al. (2003) may
certainly imply a large spread in stellar population ages, and thus an
iterative mass assembly and star formation for the central star clusters, as
discussed by the authors.  However, the nuclei that are typically selected for
the spectroscopic investigations populate the bright end of the luminosity
distribution of nuclei. Walcher et al.\ (2003) stress that in their sample
there is no indication that brighter means younger; nevertheless, it is still
possible that selection effects are important and that fainter nuclei may have
less complex mass assembly and star formation histories. A wide range of star 
formation histories would be more consistent with a process of growth of 
central star clusters that is regulated by local physics, for instance 
by the amount of fuel (either gas or smaller star clusters) available at 
various epochs in the circumnuclear regions, the angular momentum distribution 
or orbital structure of this ``fuel,'' and the physical state of the central 
regions of the disk (e.g., its density or dynamical temperature, in turn 
determining or originating from the steepness of the gravitational potential, 
the conditions to develop non-axisymmetric perturbations on small scales, 
etc.).  Furthermore, it is still unknown whether fuel-starved, silent 
AGN engines --- massive black holes --- may be present in the central star 
clusters (e.g., Marconi et al.\ 2003). The question of whether massive black 
holes reside in general in the centers of star clusters is far from settled. 
The case of G1, a globular cluster in Andromeda in which a central black hole 
of the mass expected from the linear extrapolation of the relationship 
reported for the massive spheroids (e.g., Gebhardt et al.\ 2000) has been 
detected (Gebhardt, Rich, \& Ho 2002),
argues for the presence of massive black holes in the centers of star
clusters, and supports the suggestion that black holes are ubiquitous and
proportionally sized in all spheroids, from mass scale of globular clusters
to elliptical galaxies.  A small, $\sim 10^{4-5} M_\odot$ black hole
is found embedded in the central star cluster of NGC 4395, one of the least
luminous and nearest known Type 1 Seyfert galaxies (Filippenko \& Ho 2003).
On the other hand, the nondetection of a central black hole in the central
star cluster of M33 contrasts with the G1 case and argues for the absence of
massive black holes in the centers of the distinct nuclei of bulgeless disks.
Gebhardt et al.\ (2001) discuss that, if the mass of a central black hole in
the nucleus of M33 was related to its velocity dispersion in the same way that
the known supermassive black holes are related to the dispersions of their
bulges, then a black hole with mass in the range $\sim
7\times10^3-6\times10^4\, M_\odot$ would be expected, well above the measured
upper limit of 1500 $M_\odot$.  Solutions to this inconsistency include those
suggested by the authors: the relationship between the mass of the
black hole and the velocity dispersion of the host spheroid may be nonlinear;
the conditions to make a massive black hole were better in the earlier, denser
Universe, when the stars in G1 were made; or M33's young nucleus has not
had enough time to create its own black hole.  Given the observational
uncertainties, other possibilities remain.  It could be that G1 is not a
star cluster but a harassed spheroidal galaxy [a fact mentioned by Gebhardt et
al.\ (2002) but not considered by the authors as the cause for the 
discrepancy].  Another possibility is that at least in small-sized spheroids 
such as star clusters, black holes may not be ubiquitous, or there may not 
exist a tight correlation between black hole mass and spheroid mass.  Or 
perhaps normal star clusters and the central star clusters in disk
galaxies have a different origin.

The case of M33 serves also as a smoking gun in another context.  Kormendy \&
Gebhardt (2001; see also Kormendy et al. 2003) report that the same 
correlation between the mass of the central black hole and the host luminous 
spheroid holds for galaxies with both ``normal'' and kinematically cold, 
disklike bulges (i.e., the ``pseudobulges'' discussed by Kormendy 1993). In 
contrast, M33, a pure disk galaxy with no bulge component of any sort, is 
indeed found to lack a black hole.  Kormendy \& Gebhardt (2001) conclude that 
the basic requirement for making a supermassive central black hole appears to 
be that the galaxy is capable of forming
some kind of dense, bulgelike structure, whatever its nature.
Reinterpreting this comment in the light of the bulge/dense-disk conundrum
discussed above, the results of Kormendy \& Gebhardt (2001) and Gebhardt et
al.\ (2001) may imply that the requirement for making a supermassive central
black hole is that the galaxy is capable of reaching sufficiently high central
baryonic densities. Either way, from these analyses it appears that black hole
masses are not correlated with the total gravitational potential of the disks,
and thus of the host dark matter halos.  A contrasting report, however, comes 
from Ferrarese (2002) and Baes et al.\ (2003), who claim a tight correlation
between the circular velocities of galaxies and the masses of their central
supermassive black holes, and thus an intimate link between the black holes
and the host dark matter halos.  Supermassive black holes do form in some pure 
disk systems, as shown by Filippenko \& Ho (2003) for the case of NGC 4395.
However, these authors stress that in this galaxy the estimated black hole
mass is consistent with the $M_\bullet-\sigma$ relation of Tremaine
et al.\ (2002), if the central cluster is considered in lieu of the bulge. For
a $\sigma=30$ km s$^{-1}$, a good upper limit for the velocity dispersion of
central star cluster in NGC 4395, this relation predicts a
$M_\bullet=6.6\times10^4 \, M_\odot$, consistent with the mass independently
estimated from the AGN properties (Filippenko \& Ho 2003).  Furthermore, it
remains a fact that M33, possibly the best candidate to test for the validity
of a correlation between the black hole mass and the dark matter halo mass,
appears not to support it.  As stressed by Gebhardt et al.\ (2001), if a black
hole in M33 were indeed related to the dark matter potential well, then M33
should contain a black hole of mass significantly in excess of 10$^6 M_\odot$,
which it does not.  It may be best to wait for the observational picture to be
cleared up before attempting interpretations of the claimed correlation
between black hole and dark halo masses.

Finally, given the large frequency of occurrence of nuclei in disk galaxies
and the generally accepted idea of hierarchical galaxy assembly, an
interesting question is whether the formation and evolution of the nuclei of
disk galaxies play any relevant role in the formation of supermassive black
holes in the centers of galaxies.  More generally, a key question for the
future is whether the nearly ubiquitous nuclei are a nuance or rather an
important ingredient in the formation process of disk galaxies.

\section{Concluding Remarks}

In summary, at a resolution of typically a few tens to a few hundred parsecs,
the local disk galaxy population appears to host bulges that more and more
resemble disks, and disks that more and more resemble bulges. Disks may be
denser than exponential, and bulges may be less steep than de~Vaucoluleur's
structures. Bulges are claimed to have the kinematics of thickened disks, and
buckled bars can be as dynamically hot as the structures claimed to be
``true'' bulges.  The average stellar population properties (i.e., stellar
ages and metallicities) remain the only surviving distinction between
``massive'' and ``small'' bulges and, more generally, between massive bulges
and the centers of disks.  This could certainly be an indication of different
formation processes, but could also be the result of similar processes
occurring at different epochs in the Universe (and thus naturally generating a
positive correlation between stellar densities and ages).  In my view, what is
needed at this point is a shift of the debate from the arena of morphological
classifications, where bulges and disks are distinct entities and the question
``what is the origin of bulges' is kept distinct from the question ``what is
the origin of disks,'' to one where disk galaxies are studied as a whole
without the constraints of a rigid classification scheme. The historical focus
on morphology, while highlighting many details in the trees, may in fact have
hidden the true nature of the forest. The expected outcome will be a renewed
concept of the ``Hubble sequence'' that will be ultimately be based on
physical rather than morphological considerations.  Clarifying what we really
see nearby as the endpoint of the galaxy evolution process is essential in
order to meaningfully answer how the distant progenitors, which are seen in
the most remote regions of the Universe, transform themselves to become the
descendants that populate our own surroundings.

\vspace{0.3cm}
{\bf Acknowledgements}.
I thank L. Ho for the invitation to this very stimulating meeting, and
especially for his patience waiting for this manuscript.  I am grateful to my
collaborators, V. Debattista, L.  Mayer and B.  Moore for kindly making
available some of our results prior to publication.  Many thanks to F. 
van~den~Bosch and S.  Lilly for comments on a previous version of this 
manuscript.

\end{document}